# PT-symmetric optical potentials in a coherent atomic medium

**Jiteng Sheng, [1] Mohammad-Ali Miri, [2] Demetrios N. Christodoulides, [2] and Min Xiao[1,3,*]**

[1]*Department of Physics, University of Arkansas, Fayetteville, Arkansas 72701, USA*

[2]*CREOL, College of Optics and Photonics, University of Central Florida, Orlando, Florida 32816, USA*

[3]*National Laboratory of Solid State Microstructures and Department of Physics, Nanjing University, Nanjing 210093, China*

*\*mxiao@uark.edu*

**Abstract:** We demonstrate that a coherently-prepared four-level atomic medium can provide a versatile platform for realizing parity-time (PT) symmetric optical potentials. Different types of PT-symmetric potentials are proposed by appropriately tuning the exciting optical fields and the pertinent atomic parameters. Such reconfigurable and controllable systems can open up new avenues in observing PT-related phenomena with appreciable gain/loss contrast in coherent atomic media.

Non-Hermitian parity-time (PT) symmetric Hamiltonians have attracted considerable attention since they were first proposed by Bender and Boettcher in 1998 [1]. Under certain conditions, this class of Hamiltonians displays entirely real spectra-a property often thought to belong exclusively to Hermitian systems [2]. Quite recently, this counter-intuitive behavior has been theoretically and experimentally demonstrated in a number of optical settings that judiciously engage both gain and loss processes [3-11]. This was accomplished by exploiting the mathematical isomorphism existing between the quantum Schrödinger and the paraxial wave equations. Along different lines, in multi-level atomic media, both the linear and nonlinear optics of electromagnetically induced transparency (EIT) phenomena have been intensely investigated in the past two decades [12]. In such arrangements, a host of intriguing spatial phenomena have been observed in EIT-related systems, including electromagnetically induced focusing and waveguiding [13-17], as well as self-imaging [18], grating behavior [19] and soliton propagation [20-22].

In this Letter, we show that PT-symmetric conditions can be achieved in an atomic assemble involving EIT-related four-level atoms by spatially engineering the complex refractive indices of the coherently-prepared atomic medium. In this regime, PT symmetry demands that $n(x) = n^*(-x)$, i.e., the real and imaginary parts of the refractive index profiles must be symmetric and anti-symmetric functions of the transverse coordinate x, respectively [3-11]. The realization of PT symmetry in atomic media offers several advantages over solid-state systems. For example, optical structures like coupled waveguide systems can be readily reconfigured, tuned, and effectively controlled in an atomic medium through various external parameters, i.e., frequency detunings and Rabi frequencies of the coupling and pump fields. In addition, by introducing or interfering coupling beams or by using spatial light modulators, PT-symmetric optical lattices can be readily established in the spatial domain. Recently, several PT-symmetric optical configurations have been suggested as means to realize nonreciprocal energy transport [23,24], perfect laser absorbers [25], laser amplifiers [26], etc. To date, all experimental studies carried out in PT-symmetric optics have relied on gain/loss arrangements within solid-state materials [10,11]. Therefore, the implementation of PT-symmetric potentials in atomic media will pave the way for new avenues in exploring such interesting phenomena.

A four-level N-type atomic system is considered, as shown in Fig. 1(a). The signal, coupling, and pump fields drive the atomic transitions |1>-|3>, |2>-|3>, and |1>-|4>, respectively. We note that four-level N-type systems with active Raman gain [27] are distinct from their counterparts that are based on two-photon absorption [28-30]. Throughout this work, and without any loss of generality, we focus our attention to the Rubidium N-type system with active Raman gain. By choosing appropriate parameters, the two coupling (blue) and two pump (green) beams, which propagate in the z direction, form two coupled waveguide structures, with one providing gain while the other absorption. The signal (red) beam propagates in the same direction along with the coupling and pump beams, as shown in Fig. 1(b).

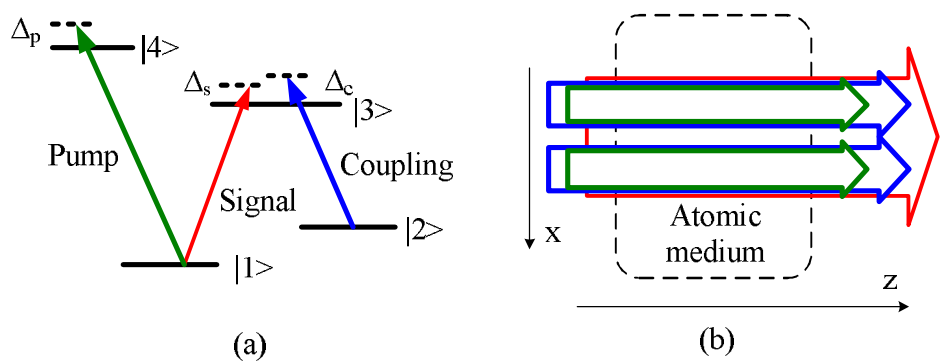

Fig. 1. (Color online) (a) The relevant four-level N-type atomic system. (b) The schematic diagram with signal (red), coupling (blue), and pump (green) fields. X and Z represent the transverse and longitudinal directions of propagation, respectively.

Under the rotating-wave approximation, the density-matrix equations for the four-level N-type atomic system are given by [27,31]:

$$\dot{\rho}_{22} = \Gamma_{42}\rho_{44} + \Gamma_{32}\rho_{33} - \Gamma_{21}\rho_{22} + \frac{i}{2}(\rho_{32}\Omega_c^* - \rho_{23}\Omega_c), \tag{1a}$$

$$\dot{\rho}_{33} = \Gamma_{43}\rho_{44} - \Gamma_{32}\rho_{33} - \Gamma_{31}\rho_{33} + \frac{i}{2}(\rho_{23}\Omega_c - \rho_{32}\Omega_c^* + \rho_{13}\Omega_s - \rho_{31}\Omega_s^*), \tag{1b}$$

$$\dot{\rho}_{44} = -(\Gamma_{43} + \Gamma_{42} + \Gamma_{41})\rho_{44} + \frac{i}{2}(\rho_{14}\Omega_p - \rho_{41}\Omega_p^*), \tag{1c}$$

$$\dot{\rho}_{21} = -\tilde{\gamma}_{21}\rho_{21} + \frac{i}{2}\rho_{31}\Omega_c^* - \frac{i}{2}\rho_{24}\Omega_p - \frac{i}{2}\rho_{23}\Omega_s, \tag{1d}$$

$$\dot{\rho}_{31} = -\tilde{\gamma}_{31}\rho_{31} + \frac{i}{2}\rho_{21}\Omega_c - \frac{i}{2}\rho_{34}\Omega_p + \frac{i}{2}(\rho_{11} - \rho_{33})\Omega_s, \tag{1e}$$

$$\dot{\rho}_{41} = -\tilde{\gamma}_{41}\rho_{41} - \frac{i}{2}\rho_{43}\Omega_s + \frac{i}{2}(\rho_{11} - \rho_{44})\Omega_p, \tag{1f}$$

$$\dot{\rho}_{32} = -\tilde{\gamma}_{32}\rho_{32} + \frac{i}{2}(\rho_{22} - \rho_{33})\Omega_c + \frac{i}{2}\rho_{12}\Omega_s, \tag{1g}$$

$$\dot{\rho}_{42} = -\tilde{\gamma}_{42}\rho_{42} - \frac{i}{2}\rho_{43}\Omega_c + \frac{i}{2}\rho_{12}\Omega_p, \tag{1h}$$

$$\dot{\rho}_{43} = -\tilde{\gamma}_{43}\rho_{43} + \frac{i}{2}\rho_{13}\Omega_p - \frac{i}{2}\rho_{42}\Omega_c^* - \frac{i}{2}\rho_{41}\Omega_s^*, \tag{1i}$$

where $\Omega_s = \mu_{13}E_s/\hbar$, $\Omega_c = \mu_{23}E_c/\hbar$, and $\Omega_p = \mu_{14}E_p/\hbar$ are the Rabi frequencies of the signal, coupling, and pump fields, respectively. For simplicity we define $\tilde{\gamma}_{21} = \gamma_{21} - i(\Delta_s - \Delta_c)$, $\tilde{\gamma}_{31} = \gamma_{31} - i\Delta_s$, $\tilde{\gamma}_{41} = \gamma_{41} - i\Delta_p$, $\tilde{\gamma}_{32} = \gamma_{32} - i\Delta_c$, $\tilde{\gamma}_{42} = \gamma_{42} - i(\Delta_c + \Delta_p - \Delta_s)$, and $\tilde{\gamma}_{43} = \gamma_{43} - i(\Delta_p - \Delta_s)$. $\Gamma_{nm}$ is the natural decay rate between level |n> and level |m>; and $\gamma_{nm} = (\Gamma_n + \Gamma_m)/2$. Here, $\Delta_s = \omega_s - \omega_{31}$, $\Delta_c = \omega_c - \omega_{32}$, and $\Delta_p = \omega_p - \omega_{41}$ are the frequency detunings for the signal, coupling, and pump fields, respectively. The susceptibility of the atomic medium can be obtained through the expression $\chi = \frac{2N\mu_{13}}{\varepsilon_0 E_s}\rho_{31}$. Given that $n = \sqrt{1+\chi} \approx 1 + \chi/2$, $\chi = \chi' + i\chi''$, and $n = n_0 + n_R + in_I$, the real and imaginary parts of the refractive index can be written as $n_R \approx \frac{1}{2}\chi' = \frac{N\mu_{13}}{\varepsilon_0 E_s}\text{Re}(\rho_{31})$ and $n_I \approx \frac{1}{2}\chi'' = \frac{N\mu_{13}}{\varepsilon_0 E_s}\text{Im}(\rho_{31})$. Here $n_0 = 1$ is the background index of the atomic medium.

We first begin by considering the realization of a dual PT-symmetric potential that could allow exchange of optical energy. This is possible by employing two different coupling fields side by side-each having an identical Gaussian intensity profile while the susceptibility of the signal field varies in the x direction as a function of the coupling field intensity. Since the coupling fields are from two independent lasers and the overlapped region between them is relatively small, the phase effect between them is not considered here. In this case the total spatial intensity distribution of the coupling beams has the form of

$$I_c(x) = A(e^{\frac{-(x-a)^2}{2\sigma^2}} + e^{\frac{-(x+a)^2}{2\sigma^2}}), \quad (2)$$

where A is a constant, 2a is the separation between the two potential channels, and $2\sqrt{2\ln 2}\sigma$ is the full-width-at-half-maximum (FWHM) of the beam width. Even though the coupling intensity profiles for these two channels are identical, by choosing different coupling frequency detunings ($\Delta_c$), one can actually introduce gain in one waveguide and absorption in the other one. The advantage offered by this scheme is that the refractive index now spatially varies only with the coupling intensity, since all the other parameters are fixed in each waveguide.

In this arrangement, perhaps the most challenging task is to find the proper coupling frequency detunings for a simultaneous gain and absorption in these two channels. To do so, we first have to identify the relation between the susceptibility and the coupling frequency detuning.

By solving Eqs. (1) numerically, one can obtain the real (dispersion) and imaginary (gain/absorption) parts of the susceptibility versus the coupling detuning for various coupling intensities as shown in Figs. 2(a) and (b), respectively. The parameters used in this example are: $\Delta_s = \Delta_p = 0$, $\Omega_s/2\pi = 0.1$ MHz, $\Omega_p/2\pi = \Gamma_{31}/2\pi = \Gamma_{32}/2\pi = \Gamma_{41}/2\pi = 3$ MHz, $\Gamma_{21} = \Gamma_{43} = \Gamma_{42} = 0$, and $\Omega_c/2\pi = 2$, 1, and 0.5 MHz for black (solid), red (dashed), and blue (dotted) curves, respectively. To achieve waveguiding, the signal field is focused at the center of each potential channel that is formed by the joint action of the coupling and pump beams, i.e., the maximum coupling intensity regions. Therefore, the real part of the susceptibility needs to get larger with increasing coupling intensity. For this reason, the coupling detunings should be negative. Also, we notice that the imaginary part of the susceptibility is close to zero when the coupling detuning is ~ -1.7155 MHz, as shown in Fig. 2(b), and in the vicinity of this zero point, absorption is induced on the left side, while gain on the right. Such property called refractive index enhancement with vanishing absorption has been widely studied in both near-resonant [32] and far-off resonant [33] atomic systems.

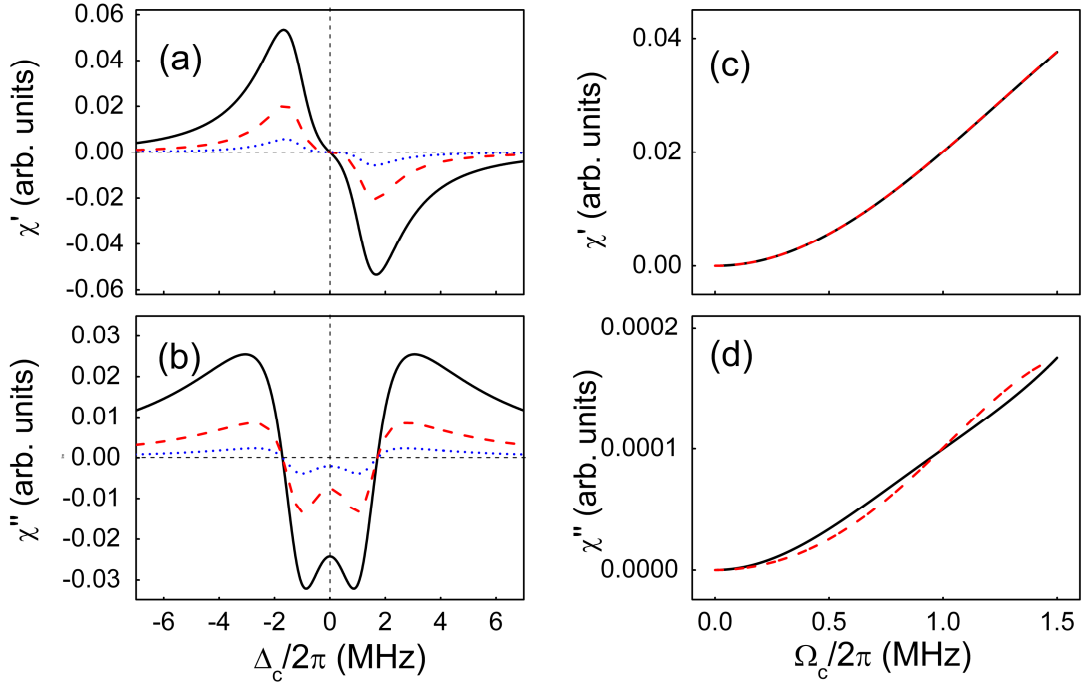

Fig. 2. (Color online) (a) The real (dispersion) and (b) imaginary (gain/absorption) parts of the susceptibility versus the coupling frequency detuning. $\Omega_c/2\pi$ = 2, 1, and 0.5 MHz for black (solid), red (dashed), and blue (dotted) curves, respectively. The (c) real and (d) imaginary parts of the susceptibility versus the coupling Rabi frequency. The black (solid) / red (dashed) curves represent $\chi'$ and $\chi''$ for the gain/absorption waveguides. The chosen parameters are $\Delta_c/2\pi$ = -1.711 and -1.720 MHz for the gain and absorption waveguides, respectively.

Next, we choose two different coupling frequency detunings so as to simultaneously introduce gain and absorption in the two separate waveguides. This implies that the only difference between these two channels is the values of coupling detuning, while all other parameters remain to be identical. To achieve this goal, we need to get the relation between the susceptibility and the coupling Rabi frequency. For example, by choosing $\Delta_c/2\pi$ = -1.711 and -1.720 MHz for the two coupling fields, we find that the real parts of the susceptibility associated with these two waveguides overlap quite well, as shown in Fig. 2(c). The imaginary parts for the gain and loss waveguides are not matched perfectly, as shown in Fig. 2(d) (in order to directly compare the two curves, the curve for the gain waveguide is flipped by multiplying a minus sign for the black solid curve). In Fig. 2(c), the black solid and red dashed curves represent the real indices for the formed gain and absorption waveguides, respectively.

Once a relation is established between the susceptibility and the Rabi frequency of the coupling field, one can then obtain the spatial index modulation. For the intensity distribution described in Eq. (2), and by choosing the FWHM ($2\sqrt{2\ln 2}\sigma$) to be 7 μm when the separation between the two waveguides (2a) is 20 μm, the intensity-dependent index graphs displayed in Figs. 2(c) and (d) can be converted into their corresponding index landscapes, as shown in Fig. 3. It is clear that the real part of the refractive index distribution is an even function of position x, whereas, at the same time, the imaginary part is odd. The value of the refractive index can be easily modified by changing the atomic density. Here, we assume the atomic density of rubidium atoms to be ~ $10^{12}$ cm$^{-3}$. In this case the maximum contrast in the real part of the refractive index is in the order of $10^{-3}$ and, thus, each optical potential channel can support only one mode. Interestingly, both the real and imaginary parts scale equally when the atomic density is changed. For realizing PT symmetry it is important that one can control the ratio between the real and imaginary indices since it determines whether the system is at below or above the PT-symmetric threshold [5]. In order to change the ratio, the two coupling frequency detunings need to be altered, so as to change the real and imaginary parts simultaneously as dictated by the Kramers-Kronig relations. For the chosen parameters, the real part changes less than the imaginary part. When the coupling detunings move away from the near zero point ($\Delta_c/2\pi$ ≈ -1.7155 MHz), the ratio between the real and imaginary indices can be modified from ~ 100 to 10. Therefore, the system can operate either below (Fig. 3(a)) or above (Fig. 3(b)) the threshold point, or can even behave as a passive one without PT symmetry (Fig. 3(c)). Here, the ratio between the real and imaginary indices is ~ 81, 10, and 3000, for Figs. 3(a-c), respectively.

Under these conditions the PT-symmetric threshold is ~ 27- as determined by the complex index profiles of the involved waveguides. It is worth to point out that the realization of transverse spatial modulation without absorption or gain as shown in Fig. 3(c) can be easily converted to periodically controllable photonic structures by using standing-waves along the wave propagation direction [34].

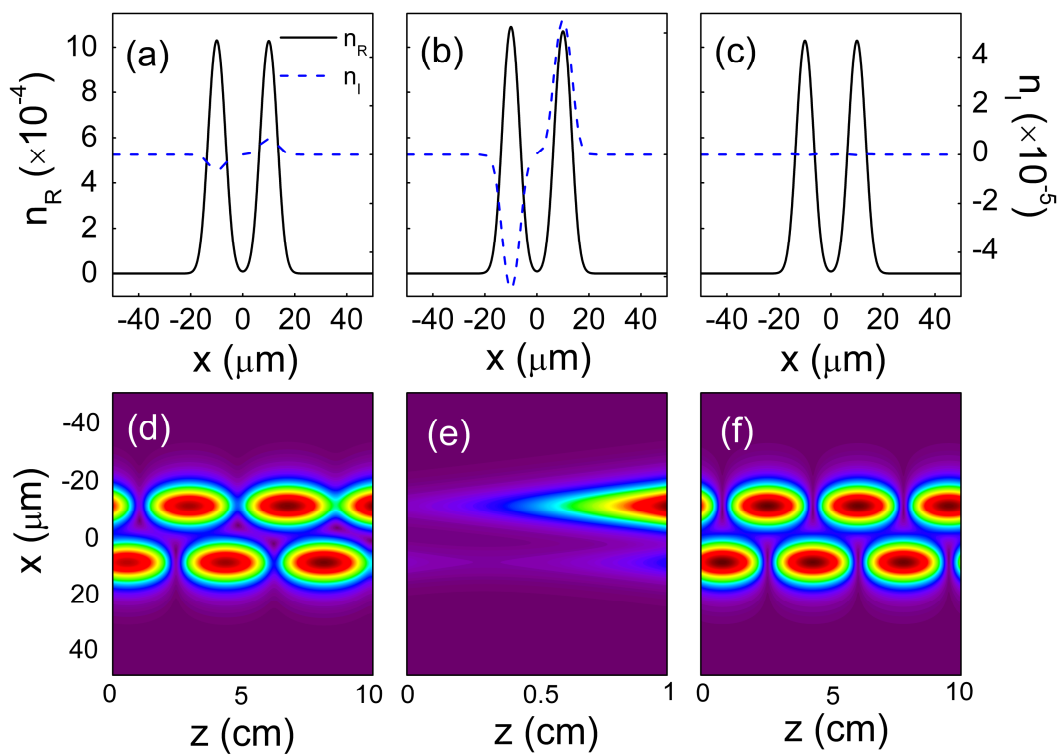

Fig. 3. (Color online) The real (solid black curves) and imaginary (dashed blue curves) refractive indices as a function of position x for (a) below threshold, (b) above threshold, and (c) conventional Hermitian cases. The parameters used are $\Delta_c/2\pi$ = -1.7100 and -1.7214 MHz for the gain and loss waveguides in part (a); $\Delta_c/2\pi$ = -1.6700 and -1.7645 MHz for waveguides in (b); and $\Delta_c/2\pi$ = -1.7155 in (c), respectively. The atomic density is ~ $10^{12}$ cm$^{-3}$ while the other parameters are the same as in Fig. 2. Pictures (d-f) show optical beam propagation patterns for the signal beam in the index potentials of (a-c), respectively.

Figures 3(d-f) show how a signal beam will propagate in the corresponding index structures as given in Figs. 3(a-c), respectively [10]. As clearly indicated by these figures, in the PT-symmetric case shown in Fig. 1, the rate at which energy is transferred crucially depends on the direction of the flow (gain to loss or loss to gain). Furthermore, by increasing the gain/loss contrast, a transition from stable to exponentially growing modes can occur, signifying the onset of PT-symmetry breaking. Note that as opposed to use Gaussian profiles, one could use Bessel-Gauss beams to overcome the severe diffraction effect in setting up this type of PT-symmetric potentials [35]. Also, either single- or multi-mode signal beam can be supported in this system, which depends on the properties of the induced waveguides, such as their widths and refractive indices.

Perfect situations are usually considered when the optical properties of PT-symmetric systems are studied, i.e., perfectly symmetric real and anti-symmetric imaginary parts of the refractive index. However, when constructing a real optical PT-symmetric potential in a practical system, such ideal conditions are difficult to reach, especially with the atomic media. Here, we define the asymmetry function as $\Delta(x) = n(x) - n^*(-x)$. Making $\Delta(x)$ close to zero is the goal when optimizing parameters. By analyzing the asymmetry function, we find that the degrees of asymmetry for the real and imaginary parts in Fig. 3 are below 5%, and therefore manageable in practical systems.

Next, we consider the cases for positive coupling detunings, so that the real index becomes smaller as the coupling intensity increases. We also introduce additional coupling beams in order to realize an optical lattice potential, as shown in Fig. 4(a), since optical lattices are more promising for realizing PT symmetry than coupled waveguides because they posses much richer phenomena and have no diffraction issue. Unlike the configuration discussed so far, each waveguide channel in this lattice case is filled with half gain and half loss. The ratio between the real and imaginary indices in Fig. 4(a) is now ~ 6.7. To achieve this, we simply move the coupling detuning farther away from the zero point ($\Delta_c/2\pi \approx 1.7155$ MHz). However, this leads to an increase in the asymmetry of the real index. To balance the real and imaginary parts, different pump intensities for the gain and absorption regimes are now necessary. The parameters used in Fig. 4(a) are $\Delta_c/2\pi = 1.65$ (1.74) MHz and $\Omega_p/2\pi = 3$ (2.943) MHz for the gain (absorption) regimes. The rest of the parameters are the same as in Fig. 3. The ratio between the imaginary and the real parts can be controlled by properly choosing the coupling detunings and the pump intensities for this case, similar to the way used in Fig. 3. The band structure of this lattice reveals an interesting property which has not been observed in other periodic lattices, i.e., by increasing the gain/loss contrast, the first band remains intact from symmetry breaking. Instead, the band merging effect starts from the higher-order bands. This is due to the fact that the gain and loss regions are mostly confined to the two sides of each waveguide channel and thus have a minimum overlap with the lowest-order Floquet-Bloch mode of the first band (that is mostly confined to the center of each channel). On the other hand, the second-band Floquet-Bloch wave functions overlap more effectively with these gain and loss regions and therefore this band is the first one to break its PT symmetry.

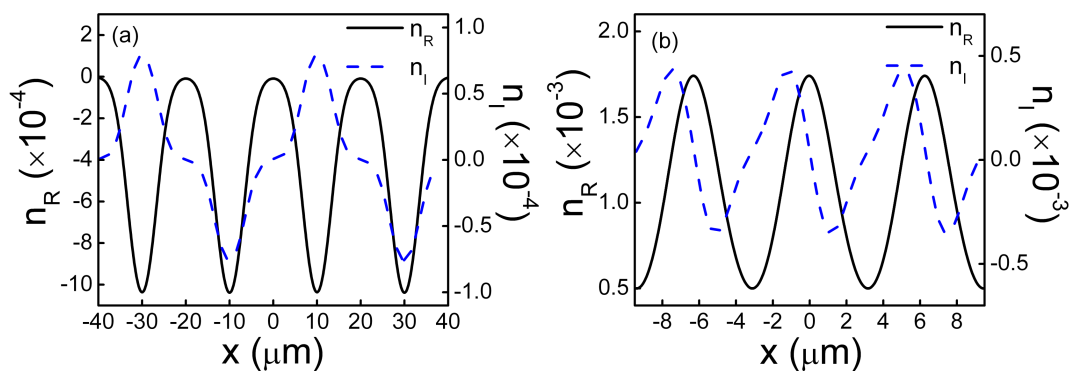

Fig. 4. (Color online) The real (solid black curves) and imaginary (dashed blue curves) parts of the refractive index for periodic lattices as a function of position x when (a) the coupling intensity and coupling detuning are spatially modified, as well as (b) the coupling and pump intensities are spatially modified.

The relation between the coupling efficiency and the separation of adjacent channels has been studied in Ref. [3]. We can also reduce the separation between the two coupled waveguides to get a stronger coupling efficiency. In such a case, the imaginary part displays a discontinuity between the induced gain and absorption regions. This is because that although the spatial variation of the coupling intensity is continuous between the gain/loss regions, the coupling frequency detuning is not, which changes independently in the gain and absorption regions to achieve the required gain and absorption values.

Instead of using continuously-varying coupling intensity and discontinuously-varying the coupling frequency detuning to generate the optical lattice potentials, an alternative way is to use continuously-varying coupling and pump intensities. For example, the coupling and pump beam intensities are spatially modified as cosine and sine functions, respectively, while the other parameters remain uniform in space. In such case, a PT-symmetric optical lattice potential can be established as shown in Fig. 4(b). The parameters used in generating Fig. 4(b) are $\Omega_c/2\pi = 1.0+0.354\cos(x)$ MHz, $\Omega_p/2\pi = 3.4+0.430\sin(x)$ MHz, $\Delta_c/2\pi = -1.701$ MHz, and $\Delta_p/2\pi = -1.100$ MHz. All the other parameters are the same as in Fig. 3. The ratio between the real and imaginary indices in Fig. 4(b) is now ~ 1.5, which can be changed to be below and above or PT-symmetry breaking threshold by modifying the amplitudes of the sine/cosine functions. The band structure of such optical lattice is also analyzed and shown to be consistent with previous studies- as in Ref. [4]. It is worth noting that if standing-wave coupling and pump fields in the z-direction have a fixed relative phase between them, then a periodic photonic structure having PT-symmetry can be implemented [36]. Moreover, the lattice potential can be extended to two-dimensions (2-D) when the coupling and pump intensities are simultaneously modulated in x and y directions. Figure 5 depicts a typical 2-D PT-symmetric lattice potential for the parameters used in Fig. 4(b).

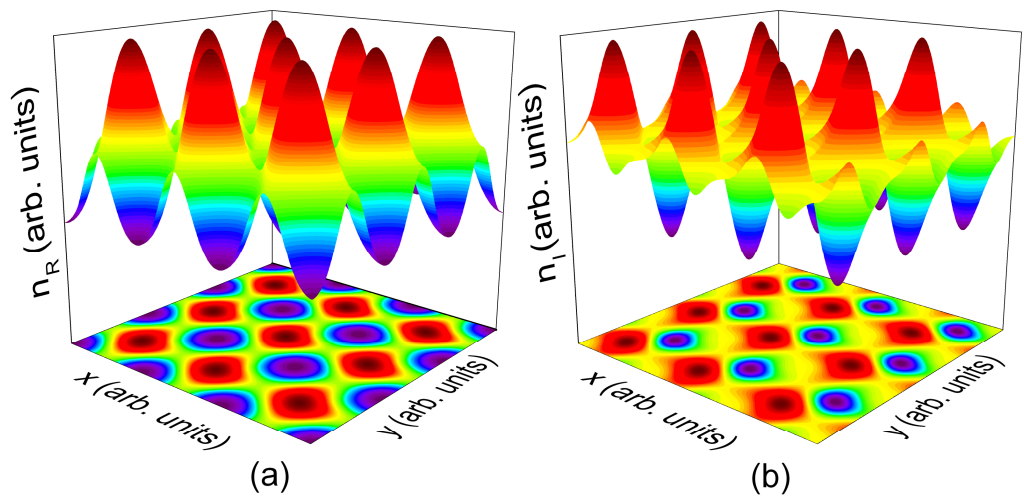

Fig. 5. (Color online) (a) Real and (b) imaginary indices for a 2-D PT-symmetric lattice potential.

By studying the wave propagation dynamics or analyzing the band structures, we find that major PT-symmetric properties can still be observed even though the generated PT-symmetric potentials for the coupled waveguides and lattices are not perfectly symmetric or anti-symmetric. To achieve a stable and accurate PT-symmetric potential, fine manipulations of the complex refractive indices are required. Since many atomic parameters can be continuously modified in the current system, the PT-symmetric potentials can always be optimized even under some non-ideal approximations, for example, when the atomic diffusion and the validity of the paraxial approximation in vapor are included.

In conclusion, we have demonstrated that a coherently-prepared four-level N-type atomic system can offer a versatile platform for exploring PT-symmetry phenomena in the optical domain. Such four-level (such as N-type) systems are quite easy to realize in real atomic media and can be easily implemented under realistic experimental conditions, in contrast to the case when a mixture of isotopes is used to provide the gain/loss mechanisms, in which very large laser powers are needed and the system can only operate below PT-symmetric threshold with a relatively restricted ratio between the real/imaginary indices [37]. In addition, in the proposed scheme, only weak light intensities are required since all laser fields operate near atomic resonances. Moreover, different types of PT-symmetric potentials have been suggested that could operate either below or above the PT-symmetric threshold, depending on relevant atomic parameters. The ability to spatially modify the complex refractive index in coherent atomic media may lead to new routes in studying and observing PT-related phenomena and processes, and find applications in quantum switching, routing, and information processing.

**Reference**

[1] C. M. Bender and S. Boettcher, Phys. Rev. Lett. **80**, 5243 (1998).

[2] C. M. Bender, Rep. Prog. Phys. **70**, 947 (2007), and references therein.

[3] R. El-Ganainy, K. G. Makris, D. N. Christodoulides, and Z. H. Musslimani, Opt. Lett. **32**, 2632 (2007).

[4] K. G. Makris, R. El-Ganainy, D. N. Christodoulides, and Z. H. Musslimani, Phys. Rev. Lett. **100**, 103904 (2008).

[5] S. Klaiman, U. Günther, and N. Moiseyev, Phys. Rev. Lett. **101**, 080402 (2008).

[6] M. C. Zheng, D. N. Christodoulides, R. Fleischmann, and T. Kottos, Phys. Rev. A **82**, 010103 (2010).

[7] Y. Chong, L. Ge, and A. Stone, Phys. Rev. Lett. **106**, 093902 (2011).

[8] A. A. Sukhorukov, Z. Xu, and Y. S. Kivshar, Phys. Rev. A **82**, 043818 (2010); A. Miroshnichenko, B. Malomed, and Y. Kivshar, Phys. Rev. A **84**, 012123 (2011).

[9] S. Longhi, Phys. Rev. Lett. **103**, 123601 (2009).

[10] A. Guo, G. J. Salamo, D. Duchesne, R. Morandotti, M. Volatier-Ravat, V. Aimez, G. A. Siviloglou, and D. N. Christodoulides, Phys. Rev. Lett. **103**, 093902 (2009) ; C. E. Rüter, K. G. Makris, R. El-Ganainy, D. N. Christodoulides, M. Segev, and D. Kip, Nat. Phys. **6**, 192 (2010).

[11] A. Regensburger, C. Bersch, M.-A. Miri, G. Onishchukov, D. N. Christodoulides, and U. Peschel, Nature **488**, 167 (2012).

[12] S. E. Harris, Phys. Today **50**, 36 (1997); H. Wang, D. Goorskey, and M. Xiao, Phys. Rev. Lett. **87**, 073601 (2001).

[13] R. R. Moseley, S. Shepherd, D. J. Fulton, B. D. Sinclair, and M. H. Dunn, Phys. Rev. Lett. **74**, 670 (1995).

[14] R. Kapoor and G. S. Agarwal, Phys. Rev. A **61**, 053818 (2000).

[15] A. G. Truscott, M. E. J. Friese, N. R. Heckenberg, and H. Rubinsztein-Dunlop, Phys. Rev. Lett. **82**, 1438 (1999).

[16] P. K. Vudyasetu, D. J. Starling, and J. C. Howell, Phys. Rev. Lett. **102**, 123602 (2009).

[17] O. Firstenberg, P. London, M. Shuker, A. Ron, and N. Davidson, Nat. Phys. **5**, 665 (2009).

[18] J. Cheng and S. Han, Opt. Lett. **32**, 1162 (2007).

[19] H. Ling, Y. Li, and M. Xiao, Phys. Rev. A **57**, 1338 (1998).

[20] T. Hong, Phys. Rev. Lett. **90**, 183901 (2003); Y. Wu and L. Deng, Phys. Rev. Lett. **93**. 143904 (2004).

[21] H. Michinel, M. J. Paz-Alonso, and V. M. Pérez-García, Phys. Rev. Lett. **96**, 023903 (2006).

[22] Y. P. Zhang, Z. Wang, Z. Nie, C. Li, H. Chen, K. Lu, and M. Xiao, Phys. Rev. Lett. **106**, 093904 (2011).

[23] H. Ramezani, T. Kottos, R. El-Ganainy and D. N. Christodoulides, Phys. Rev. A **82**, 043803 (2010).

[24] Z. Lin, H. Ramezani, T. Eichelkraut, T. Kottos, H. Cao, and D. N. Christodoulides, Phys. Rev. Lett. **106**, 213901 (2011).

[25] S. Longhi, Phys. Rev. A **82**, 031801(R) (2010).

[26] M.-A. Miri, P. LiKamWa, and D. N. Christodoulides, Opt. Lett. **37**, 764 (2012).

[27] H. Kang, L. Wen, and Y. Zhu, Phys. Rev. A **68**, 063806 (2003).

[28] H. Schmidt and A. Imamogdlu, Opt. Lett. **21**, 1936 (1996).

[29] S. E. Harris and Y. Yamamoto, Phys. Rev. Lett. **81**, 3611 (1998).

[30] J. Sheng, X. Yang, U. Khadka, and M. Xiao, Opt. Express **19**, 17059 (2011); J. Sheng, X. Yang, H. Wu, and M. Xiao, Phys. Rev. A **84,** 053820 (2011).

[31] M. O. Scully and M. S. Zubairy, Quantum Optics (Cambridge University Press, Cambridge, England, 1997).

[32] M. O. Scully, Phys. Rev. Lett. **67**, 1855 (1991); M. Fleischhauer, C. H. Keitel, M. O. Scully, C. Su, B. T. Ulrich, and S. Y. Zhu, Phys. Rev. A **46**, 1468 (1992); A. S. Zibrov, M. D. Lukin, L. Hollberg, D. E. Nikonov, M. O. Scully, H. G. Robinson, and V. L. Velichansky, Phys. Rev. Lett. **76**, 3935 (1996).

[33] D. D. Yavuz, Phys. Rev. Lett. **95**, 223601 (2005); N. A. Proite, B. E. Unks, J. T. Green, and D. D. Yavuz, Phys. Rev. Lett. **101**, 147401 (2008).

[34] C. O'Brien and O. Kocharovskaya, Phys. Rev. Lett. **107**, 137401 (2011).

[35] J. Durnin, J. J. Miceli, Jr.*,* J. H. Eberly, Phys. Rev. Lett. **58**, 1499 (1987).

[36] L. Feng, Y. Xu, W. S. Fegadolli, M. Lu, J. E. B. Oliveira, V. R. Almeida, Y. Chen, and A. Scherer, Nature Materials, **12**, 108 (2013).

[37] C. Hang, G. Huang, and V. V. Konotop, Phys. Rev. Lett. **110**, 083604 (2013).